\documentstyle[11pt,newpasp,twoside,epsf]{article}
\newcommand\Mdot  {\dot{M}}
\newcommand\Mwd   {M_{\rm wd}}
\newcommand\Rwd   {R_{\rm wd}}

\newcommand\lax{{\lower0.75ex\hbox{ $<$       }\atop\raise0.5ex\hbox{ $\sim$ }}}
\newcommand\gax{{\lower0.75ex\hbox{ $>$       }\atop\raise0.5ex\hbox{ $\sim$ }}}
\newcommand\pax{{\lower0.75ex\hbox{ $\propto$ }\atop\raise0.5ex\hbox{ $\sim$ }}}

\def\edcomment#1{\iffalse\marginpar{\raggedright\sl#1\/}\else\relax\fi}
\reversemarginpar

\begin{document}

\title{Plasma Diagnostics in High-Resolution X-ray Spectra
       of Magnetic Cataclysmic Variables}
 \author{Christopher W.~Mauche}
\affil{Lawrence Livermore National Laboratory,
       L-43, 7000 East Avenue, Livermore, CA 94550}

\begin{abstract}
Using the {\it Chandra\/} HETG spectrum of EX~Hya as an example, we
discuss some of the plasma diagnostics available in high-resolution X-ray
spectra of magnetic cataclysmic variables. Specifically, for conditions
appropriate to collisional ionization equilibrium plasmas, we discuss the
temperature dependence of the H- to He-like line intensity ratios and the
density and photoexcitation dependence of the He-like $R$ line ratios and
the Fe~XVII $I(17.10~{\rm \AA})/I(17.05~{\rm \AA })$ line ratio. We show
that the plasma temperature in EX~Hya spans the range from $\approx 0.5$
to $\approx 10$ keV and that the plasma density $n\gax 2\times 10^{14}~\rm
cm^{-3}$, orders of magnitude greater than that observed in the Sun or
other late-type stars.
\end{abstract}

%%%%%%%%%%%%%%%%%%%%%%%%%%%%%%%%%%%%%%%%%%%%%%%%%%%%%%%%%%%%%%%%%%%%%%%%

\section{Introduction}

In magnetic cataclysmic variables (mCVs), the flow of material lost by
the secondary is channeled onto small spots on the white dwarf surface
in the vicinity of the magnetic poles. Because the infall velocity is
supersonic, the flow passes through a strong shock above the stellar
surface, where 15/16 of its prodigious kinetic energy is converted into
thermal energy. The post-shock plasma is hydrostatically supported,
and cools via cyclotron, thermal bremsstrahlung, and line emission
before settling onto the white dwarf surface. Consequently, the 
post-shock plasma of mCVs is both multi-temperature and multi-density,
with a post-shock temperature $T\le T_{\rm shock}= 3G\Mwd \mu m_{\rm H}/
8k\Rwd\approx 250~{\rm MK}\approx 20$ keV and density $n\ge n_{\rm shock}
=\Mdot/4\pi f\Rwd ^2\mu m_{\rm H} (v_{\rm ff}/4)\approx 10^{13}~\rm
cm^{-3}$ for a mass-accretion rate $\Mdot=10^{15}~\rm g~s^{-1}$ (hence
$L=G\Mwd\Mdot/\Rwd\approx 9\times 10^{31}~\rm erg~s^{-1}$), relative
spot size $f=0.1$, and free-fall velocity $v_{\rm ff}=(2G\Mwd/\Rwd)^{1/2}
\approx 4300~\rm km~s^{-1}$. If the shock height is a small fraction
of the white dwarf radius, a significant fraction of the resulting
\hbox{X-ray} emission is intercepted by the white dwarf surface. 
Competition between photoelectric absorption and Thompson and Compton
scattering results in $\sim 20$ keV X-rays being predominantly scattered
by the white dwarf photosphere, resulting in a hard reflection spectral
component, while softer and harder X-rays are increasingly likely to
deposit their energy in the white dwarf photosphere, heating it to a
temperature $T_{\rm bb}\approx (G\Mwd\Mdot /8\pi\sigma f\Rwd ^3)^{1/4}$
$\approx 30$~kK. $\Mdot/f$ must be larger by a factor of $\sim 10^4$ to
explain the soft X-ray spectral component of AM Her-type mCVs.

The {\it unique\/} aspect of the X-ray--emitting plasma in mCVs is the
high densities, which are the result of the magnetic funneling of the
mass lost by the secondary, the factor-of-four density jump across the
accretion shock, and the settling nature of the post-shock flow, wherein
the density $n\pax T^{-1}$. X-ray line intensity ratios have long been used
to diagnose the temperature and density of solar plasma, but until recently
the effective area and spectral resolution of X-ray observatories have
been inadequate to allow similar studies of cosmic X-ray sources. The
multi-temperature nature of the plasma in CVs is readily apparent from the
broad-band nature of their X-ray continua, but even with the 0.5--10 keV
bandpass and $\Delta E=120$ eV spectral resolution of the {\it ASCA\/}
SIS detectors, it is possible to adequately describe the observed spectra
of CVs with a small number (typically 2) of discrete temperature
components. Ishida, Mukai, \& Osborne (1994) and subsequently Fujimoto
\& Ishida (1997) first applied the ratio of the intensities of the H- to
He-like lines of Mg, Si, S, Ar, and Fe in the {\it ASCA\/} SIS spectrum
of the intermediate polar-type mCV EX~Hya to constrain the range of
temperatures present in its accretion column. Unfortunately, {\it ASCA\/}
did not have the spectral resolution necessary to utilize any of the {\it
density\/} diagnostics available in the X-ray bandpass, such as the
density-sensitive He-like triplets, which require $\Delta E\lax 60$ eV.
With the launch of the {\it Chandra\/} and {\it XMM-Newton\/} X-ray
observatories, it is now possible to utilize a broad range of spectral
diagnostics to characterize the plasma of mCVs.

%Fig1%%%%%%%%%%%%%%%%%%%%%%%%%%%%%%%%%%%%%%%%%%%%%%%%%%%%%%%%%%%%%%%%%%%
\begin{figure}
\plotone{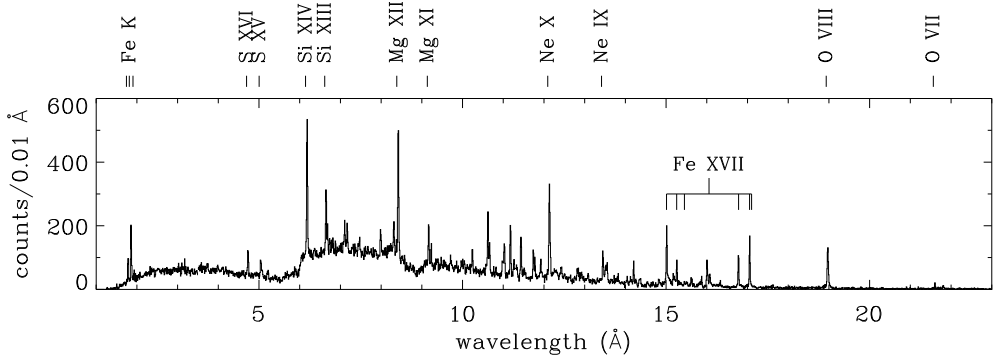}
\caption{{\it Chandra\/} HEG ($\lambda<3.5$~\AA) and MEG ($\lambda\ge
3.5$~\AA) count spectrum of EX~Hya. Identifications are provided for
emission lines of H- and He-like ions of O, Ne, Mg, Si, S, and Fe. Most
other lines are due to Fe L-shell ions; those from Fe~XVII are labeled.}
\end{figure}
%%%%%%%%%%%%%%%%%%%%%%%%%%%%%%%%%%%%%%%%%%%%%%%%%%%%%%%%%%%%%%%%%%%%%%%%

%Fig2%%%%%%%%%%%%%%%%%%%%%%%%%%%%%%%%%%%%%%%%%%%%%%%%%%%%%%%%%%%%%%%%%%%
\begin{figure}
\plotone{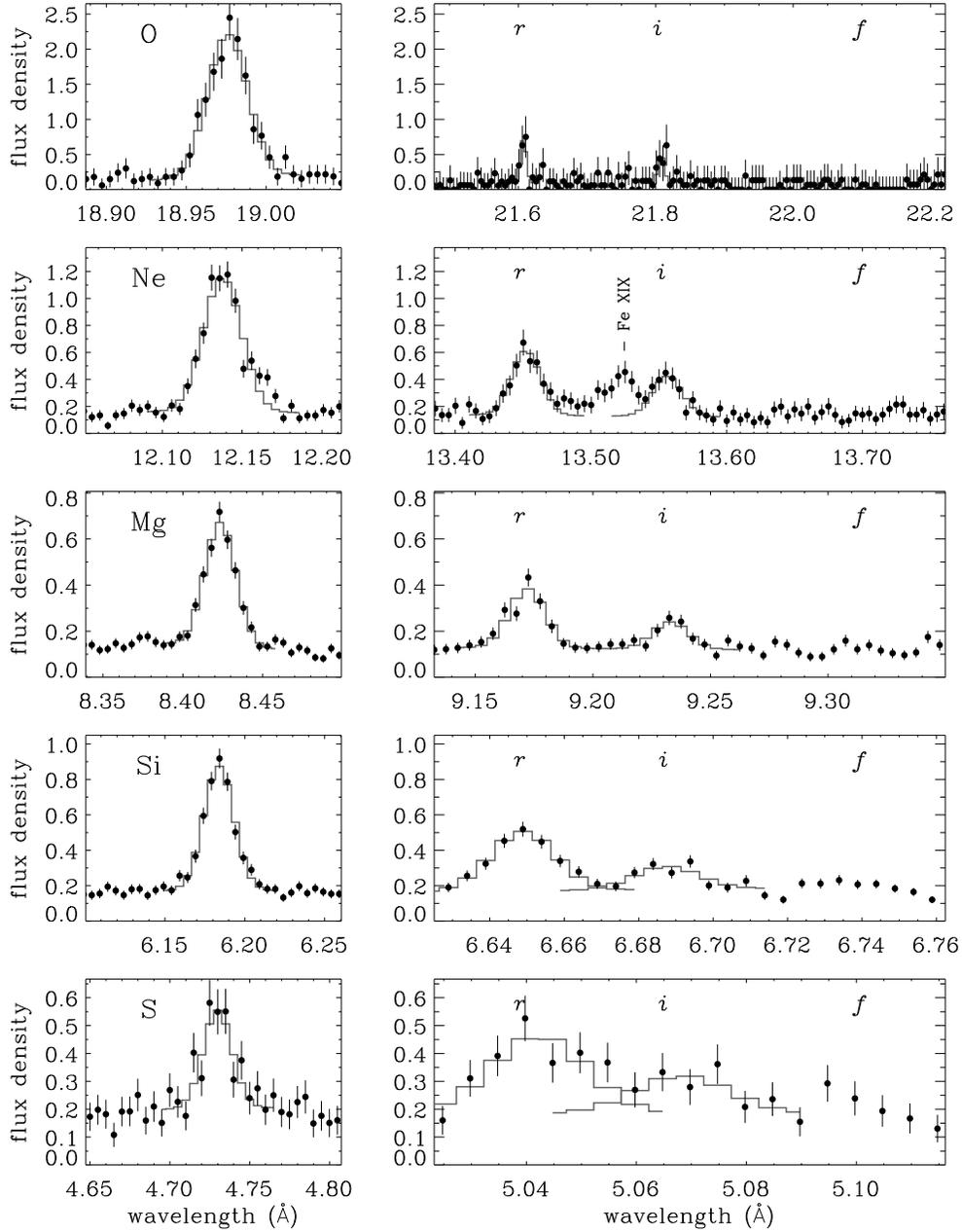}
\caption{Detailed views of the H-like Ly$\alpha $ and He-like triplet
lines of O, Ne, Mg, Si, and S in the MEG spectrum of EX~Hya. Units of flux
density are $10^{-2}~\rm photons~cm^{-2}~s^{-1}~\AA ^{-1}$. Data are shown
by the filled circles with error bars, Gaussian fits (broadened by the
spectrometer resolution) are shown by the gray histograms. Positions of
the resonance ({\it r\/}), intercombination ({\it i\/}), and forbidden
({\it f\/}) lines of the He-like triplets are indicated. Note the absence
of the forbidden lines, indicating high electron densities and/or
photoexcitation.}
\end{figure}
%%%%%%%%%%%%%%%%%%%%%%%%%%%%%%%%%%%%%%%%%%%%%%%%%%%%%%%%%%%%%%%%%%%%%%%%

\section{Chandra HETGS Observation of EX Hya}

Motivated by the beautiful {\it ASCA\/} spectrum, we recently obtained
{\it Chandra\/} HETGS ($\Delta\lambda = 0.01$ and 0.02~\AA ) and
{\it XMM-Newton\/} RGS ($\Delta\lambda = 0.07$~\AA ) and EPIC spectra of
EX~Hya. The 60 ks {\it Chandra\/} observation was performed on 2000 May
18 during a multiwavelength ({\it RXTE\/}, {\it Chandra\/}, {\it EUVE\/},
{\it FUSE\/}, {\it HST\/}, and ground-based optical) campaign, while the
{\it XMM-Newton\/} observation (100 ks) was performed on 2000 July 1--2.
Because of space limitations, we discuss here only the {\it Chandra\/}
HETG spectrum, which is shown in Figure~1. The spectrum consists of a
modest continuum with superposed lines of H- and He-like O, Ne, Mg, Si,
S, and Fe; Fe XVII--XXIV; and ``neutral'' Fe (via the weak fluor\-escent
line at 6.4 keV = 1.94~\AA ). Whereas the optical--FUV emission lines
of EX~Hya are {\it broad\/}, with $\rm FWHM \approx 7$--$10~\rm \AA \sim
2000$--$3000~\rm km~s^{-1}$ (Hellier et al.\ 1987; Greeley et al.\ 1997;
Mauche 1999), the X-ray emission lines are {\it narrow\/}, with $\rm FWHM
\approx 20~m\AA \sim 600~km~s^{-1}$. This is explained qualitatively by
the fact that (1) the post-shock velocity $v\le v_{\rm shock}=v_{\rm ff}
/4\approx 900~\rm km~s^{-1}$ and (2) we view the EX Hya binary nearly
edge-on, so the velocity vector of the post-shock flow lies more nearly on
the plane of the sky.

\section{Temperature Diagnostic}

As was done with the {\it ASCA\/} spectra, we can use the ratio of the
H- to He-like line intensities in the HETG spectrum of EX~Hya to
constrain the temperature distribution in the post-shock plasma. This
was accomplished by fitting Gaussians to the various emission lines,
accounting for the effective area and resolution function of the
spectrometer. Results of these fits are shown in Figure~2 for O, Ne, Mg,
Si, and S. In addition to easily resolving the H- and He-like emission
lines, the HETGS cleanly resolves the resonance ($r$), intercombination
($i$), and forbidden ($f$) lines of the He-like triplets, as well as
accounts for the presence of other emission lines, such as the 
relatively strong Fe~XIX emission line situated between the resonance
and intercombination lines of Ne~IX (these three lines appear as a blob
in the {\it XMM\/} RGS spectrum). Assuming the Mewe, Gronenschild,
\& van den Oord (1985) collisional ionization equilibrium (CIE) ionization
balance and line intensities, Figure~3 shows that the measured H- to
He-like line intensity ratios require that the plasma temperature extends
from $kT_{\rm min}\approx 0.5$ keV to $kT_{\rm max}\approx 10$ keV. This
is the same range inferred by Fujimoto \& Ishida (1997) from the {\it
ASCA\/} spectrum of EX~Hya, even though the SIS could not resolve the H-
and He-like emission lines of O and Ne. The lower temperature limit is a
bit of a puzzle, because it seems inescapable that the plasma will cool
below this temperature. A possible explanation is that the optical depths
in the lines begins to be important at this temperature, where the plasma
density is expected to be a factor of $\sim T_{\rm max}/T_{\rm min}\sim
20$ times higher than the value just below the accretion shock.

%Fig3%%%%%%%%%%%%%%%%%%%%%%%%%%%%%%%%%%%%%%%%%%%%%%%%%%%%%%%%%%%%%%%%%%%
\begin{figure}
\plotone{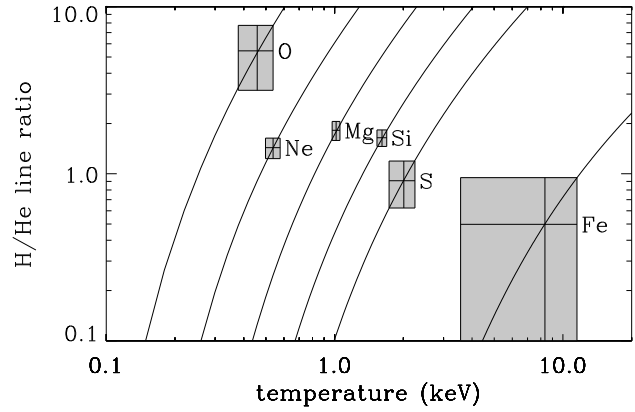}
\caption{Temperature distribution of the plasma in EX~Hya based on the
measured H- to He-like line intensity ratios. Curves assume the Mewe,
Gronenschild, \& van den Oord (1985) CIE ionization balance and line
intensities, and the error boxes are $1\, \sigma $ based on counting
statistics alone. A broad range of temperatures is indicated, from
$kT_{\rm min}\approx 0.5$ keV (probably the temperature where the optical
depth in the lines becomes important) to $kT_{\rm max} \approx 10$~keV
(comparable to the shock temperature $T_{\rm shock} = 15.4^{+5.3}_{-2.6}$
keV inferred by Fujimoto \& Ishida 1997 from the {\it ASCA\/} SIS spectrum
of EX~Hya).}
\end{figure}
%%%%%%%%%%%%%%%%%%%%%%%%%%%%%%%%%%%%%%%%%%%%%%%%%%%%%%%%%%%%%%%%%%%%%%%%

%Fig4%%%%%%%%%%%%%%%%%%%%%%%%%%%%%%%%%%%%%%%%%%%%%%%%%%%%%%%%%%%%%%%%%%%
\begin{figure}
\plotone{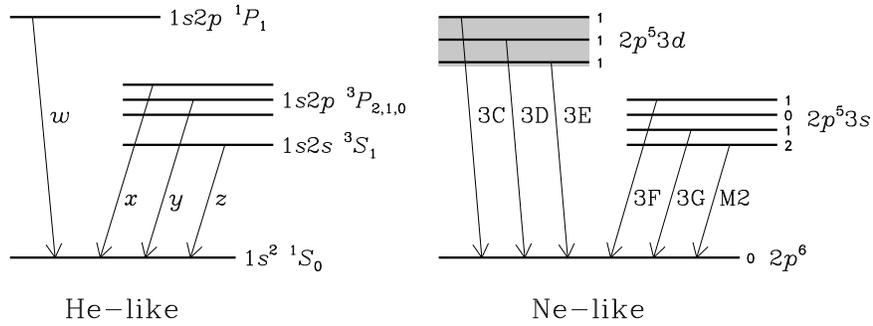}
\caption{Simplified Grotrian diagrams for He- and Ne-like ions. For the
He-like diagram the resonance ($w=r$), intercombination ($x+y=i$) and
forbidden ($z=f$) lines are labeled. For the Ne-like diagram the 3C--M2
lines are labeled; for Fe~XVII these correspond to the $\lambda = 15.01$,
15.26, 15.45, 16.78, 17.05, and 17.10~\AA \ lines, respectively, indicated
in Fig.~1.}
\end{figure}
%%%%%%%%%%%%%%%%%%%%%%%%%%%%%%%%%%%%%%%%%%%%%%%%%%%%%%%%%%%%%%%%%%%%%%%%

\section{Density Diagnostics}

\subsection{He-like R Line Ratios}

The standard density diagnostic of high-temperature plasmas is the ratio
of the forbidden to intercombination lines of He-like ions (Gabriel \&
Jordan 1969; Blumenthal, Drake, \& Tucker 1972; Porquet et al.\ 2001).
Because the $1s2s\> ^3S_1$ upper level of the forbidden line is
metastable, it can be depopulated by collisional excitation, leading to
the conversion of the $1s^2\> ^1S_0$--$1s2s\> ^3S_1$     forbidden        
line  $f$ into the    $1s^2\> ^1S_0$--$1s2p\> ^3P_{2,1}$ intercombination
blend $i$ (for the Grotrian diagram, refer to the left panel of Fig.~4).
We used the LXSS plasma code being developed at LLNL (Mauche, Liedahl, \&
Fournier 2001) to calculate the $R\equiv f/i$ line ratios as a function
of electron density for the abundant elements, and present the results in
Figure~5. It can be seen that the critical density for this ratio scales
with $Z$, ranging from $n_{\rm c}\approx 6\times 10^8~\rm cm^{-3}$ for C
to $n_{\rm c}\approx 3\times 10^{17}~\rm cm^{-3}$ for Fe. Unfortunately,
this trend is opposite to that in the accretion column of mCVs (where the
high-$Z$ ions dominate at the top of the column where the temperature is
highest and the density is lowest, and the low-$Z$ ions dominate at the
bottom of the column where the temperature is lowest and the density is
highest), limiting our ability to ``map'' the density structure of the
column. As is evident from Figure~2, all the He-like triplets of EX~Hya
are in the ``high-density limit'' ($R\approx 0$), implying that the
plasma density $n_{\rm e}\gax 10^{15}~\rm cm^{-3}$.

\subsection{Complications from Photoexcitation}

Unfortunately, this result is not entirely secure. In UV-bright sources
like early-type stars, X-ray binaries, and CVs, photoexcitation as well
as collisional excitation acts to depopulate the upper level of the
He-like forbidden lines. If the radiation field is sufficiently strong
at the appropriate wavelengths, the $R$ line ratio can be in the
``high-density limit'' regardless of the density. Photoexcitation has
been shown to explain the low $R$ line ratios of early-type stars (Kahn
et al.\ 2001; Waldron \& Cassinelli 2001), and could explain the low $R$
line ratios of EX~Hya. To investigate this effect, we included in the
LXSS level-population kinetics calculation the photoexcitation rates $(\pi
e^2/m_ec) f_{ij} F_\nu (T)$, where $F_\nu (T)$ is the continuum spectral
energy distribution and $f_{ij}$ are the oscillator strengths of the
various transitions. For simplicity, we assume that $F_\nu (T) = (4\pi
/h\nu ) B_\nu (T_{\rm bb})$ (i.e., the radiation field is that of a
blackbody of temperature $T_{\rm bb}$) and that the dilution factor of the
radiation field is equal to $1\over 2$ (i.e., the X-ray--emitting plasma
is in close proximity to the source of the photoexcitation continuum).
For $T_{\rm bb}=30$~kK, we obtain the $R$ line ratios shown by the full
curves in the right panel of Figure~5, which demonstrates that, for the
given assumptions, all of the He-like $R$ line ratios through Si~XIII are
significantly affected by photoexcitation. It is unfortunate that this
effect begins to disappear at S~XV, where the HETGS effective area is
low and the spectral resolution is only just sufficient to resolve the
He-like triplet. Calorimeter-type detectors like those planned for {\it
Astro-E2\/} and {\it Constellation-X\/} are required to work at these
short  wavelengths/high energies.

%Fig5%%%%%%%%%%%%%%%%%%%%%%%%%%%%%%%%%%%%%%%%%%%%%%%%%%%%%%%%%%%%%%%%%%%
\begin{figure}
\plotone{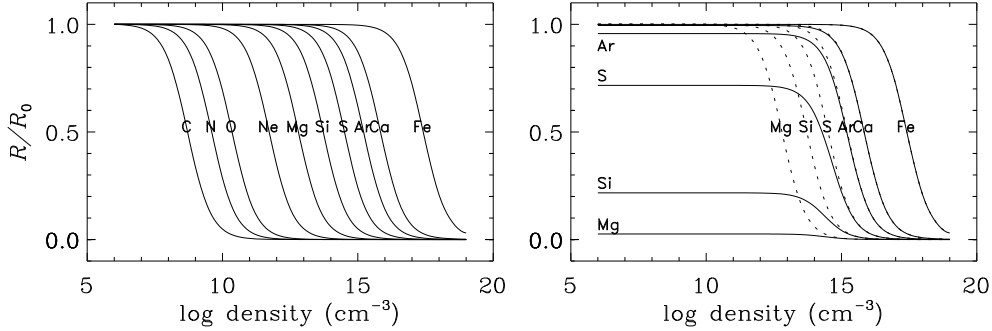}
\caption{{\it Left panel\/}: He-like $R\equiv z/(x+y)=f/i$ line ratios
of the abundant elements as a function of electron density. For C, N,
O, \ldots , Fe, the plasma temperature $T_{\rm e}= 0.46$, 0.57, 0.86,
\ldots, 36.1 MK, the temperature of the peak ionization fraction for each
He-like ion. {\it Right panel\/}: Similar to the left panel ({\it dotted
curves\/}),  but accounts for photoexcitation by a $T_{\rm bb}=30$ kK
blackbody ({\it full curves\/}). The $R$ line ratios of all ions though
Mg are in the ``high-density limit'' regardless of the density. In both
panels the line ratios are scaled to the low-density values $R_0$ for
ease of comparison.}
\end{figure}
%%%%%%%%%%%%%%%%%%%%%%%%%%%%%%%%%%%%%%%%%%%%%%%%%%%%%%%%%%%%%%%%%%%%%%%%

%Fig6%%%%%%%%%%%%%%%%%%%%%%%%%%%%%%%%%%%%%%%%%%%%%%%%%%%%%%%%%%%%%%%%%%%
\begin{figure}
\plotone{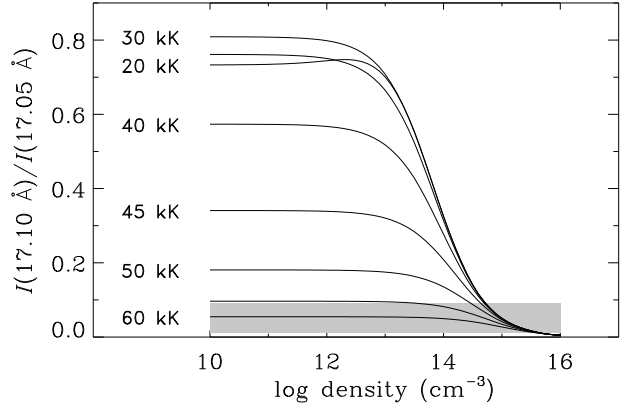}
\caption{Fe~XVII $I(17.10~{\rm \AA })/I(17.05~{\rm \AA })$ line ratio as a
function of electron density for a $T_{\rm e}= 4$~MK plasma photoexcited
by a $T_{\rm bb}= 20$, 30, 35, 40, 45, 50, 55, and 60~kK blackbody. The
gray stripe delineating the $1\,\sigma $ envelope of the line ratio
measured in EX~Hya requires that the electron density $n_{\rm e}\ga 3
\times 10^{14}~\rm cm^{-3}$ and/or the blackbody temperature $T_{\rm bb}
\ga 55$~kK.}
\end{figure}
%%%%%%%%%%%%%%%%%%%%%%%%%%%%%%%%%%%%%%%%%%%%%%%%%%%%%%%%%%%%%%%%%%%%%%%%

\subsection{Fe~XVII I(17.10~\AA )/I(17.05~\AA ) Line Ratio}

Other than the H- and He-like lines of the abundant elements, the
brightest emission lines in the {\it Chandra\/} HETG spectrum of EX~Hya
are the $2p^6$--$2p^53l$ ($l =s$, $d$) lines of Fe~XVII at 15--17~\AA \
(Fig.~1). These lines are strong in the X-ray spectra of high-temperature
CIE plasmas because of the high abundance of Fe, the persistence of
Ne-like Fe over a broad temperature range ($T_{\rm e}\approx 2$--12 MK),
and the large collision strengths for $2p\to nd$ transitions. The
importance of Fe~XVII has engendered numerous studies of its atomic
structure and level-population kinetics, and while emphasis is usually
placed on the {\it temperature\/} dependence of the $2p$--$3l$ line ratios
(e.g., Rugge \& McKenzie 1985; Smith et al.\ 1985; Raymond \& Smith 1986),
Mauche, Liedahl, \& Fournier (2001) recently discussed the {\it density\/}
dependence of these ratios.

The Grotrian diagram for Fe~XVII is shown in the right panel of Figure~4.
Like the $1s2s\> ^3S_1$ upper level of the forbidden line of He-like ions,
the $2p^53s\>(J=2)$ upper level of the 17.10~\AA \ line of Fe~XVII is
metastable, so collisional depopulation sets in at lower densities, and
the intensity ratio of the 17.10~\AA\ line to any of the other $2p$--$3l$
lines (say, the 17.05~\AA \ line) provides a diagnostic of the plasma
density. The Fe~XVII $I(17.10~{\rm \AA })/I(17.05~{\rm \AA })$ density
diagnostic is ideal for mCVs for two reasons. First, the critical density
is high: $n_{\rm c}\approx 3\times 10^{13}~\rm cm^{-3}$, comparable to
that of Si~XIII. Second, the $I(17.10~{\rm \AA })/I(17.05~{\rm \AA })$
line ratio is less sensitive than the He-like $R$ line ratios to
photoexcitation. In Fe~XVII, photoexcitations out of the $2p^53s\>(J=2)$
level go primarily into the $2p^53p$ manifold, requiring a significant
flux of photons in the 190--410~\AA \ waveband. In He-like ions,
photoexcitations out of the $1s2s\> ^3S_1$ level go primarily into the
$1s2p\> ^3P_{2,1}$ levels, requiring a significant flux of photons at
$\lambda =1623$, 1263, 1036, 865, 743, 637, 551, and 404~\AA \ for O,
Ne, Mg, Si, S, Ar, Ca, and Fe, respectively. Because the photoexcitation
continuum is typically stronger in the UV--FUV than in the EUV, the
Fe~XVII $I(17.10~{\rm \AA })/I(17.05~{\rm \AA })$ line ratio is less
sensitive to photoexcitation.

To investigate the density, temperature, and photoexcitation sensitivity
of the Fe~XVII $I(17.10~{\rm \AA }/I(17.05~{\rm \AA })$ line ratio, we
calculated LXSS atomic models of Fe~XVII for a range of densities $n_{\rm
e}=10^{10}$--$10^{16}~\rm cm^{-3}$, temperatures $T_{\rm e} =2$--8~MK
(spanning the range for which the Fe~XVII ionization fraction is $\gax
0.1$), and blackbody photoexcitation temperatures $T_{\rm bb}=20$--60~kK.
The $I(17.10~{\rm \AA })/I(17.05~{\rm~\AA })$ line ratio for $T_{\rm e}
=4$~MK, the peak of the Fe~XVII ionization fraction, is shown in Figure
6. The measured Fe~XVII $I(17.10~{\rm \AA }/I(17.05~{\rm \AA })$ line
ratio of $0.05\pm 0.04$ can be explained if the plasma density $n_{\rm e}
\gax 3\times 10^{14}~\rm cm^{-3}$ or if the photoexcitation temperature
$T_{\rm bb}\gax 55$~kK. As detailed by Mauche, Liedahl, \& Fournier
(2001), the second option is consistent with the assumptions (blackbody
emitter, dilution factor equal to $1\over 2$) and the observed FUV flux
density only if the fractional emitting area of the accretion spot $f\le
2\%$. This constraint and the observed X-ray flux requires a density
$n\gax 2\times 10^{14}~\rm cm^{-3}$ for the post-shock flow. Either way,
then, the {\it Chandra\/} HETG spectrum of EX~Hya requires that the
plasma density in this mCV is orders of magnitude greater than that
observed in the Sun or other late-type stars.

\section{Summary}

Using the {\it Chandra\/} HETG spectrum of EX~Hya as an example, we have
discussed some of the plasma diagnostics available in high-resolution
X-ray spectra of mCVs. Specifically, we have discussed the temperature
dependence of the H- to He-like line intensity ratios and the density and
photoexcitation dependence of the He-like $R$ line ratios and the Fe~XVII
$I(17.10~{\rm \AA})/I(17.05~{\rm \AA })$ line ratio. Since the discussion
assumes that the plasma is in collisional ionization equilibrium and
that the optical depths in the lines are negligible, it does {\it not\/}
apply to (1) the pre-shock flow, where photoionization competes with
collisional ionization to determine the physical state of the plasma, (2)
the immediate post-shock flow, where the ions and electrons are not in
thermal equilibrium, or (3) the very base of the post-shock flow, where
the line optical depths are non-negligible. Where the plasma is optically
thin and in collisional ionization equilibrium, the plasma temperature
spans the range from $\approx 0.5$ to $\approx 10$ keV and the plasma
density $n\gax 2\times 10^{14}~\rm cm^{-3}$. The lower temperature 
probably signals where the plasma becomes optically thick in the lines,
while the higher temperature is comparable to the shock temperature, which
Fujimoto \& Ishida (1997) inferred from the {\it ASCA\/} SIS spectrum of
EX~Hya is $T_{\rm shock} =15.4^{+5.3}_{-2.6}$ keV. 

This communication has only scratched the surface of the many plasma
diagnostics available in high-resolution X-ray spectra of mCVs. For lack
of space, our discussion has concentrated on density diagnostics and the
complicating affects of photoexcitation because these features are unique
to mCVs. For a more general discussion of the physics of high-temperatures
plasmas, the reader is referred to the volume ``X-ray Spectroscopy in
Astrophysics'' (van Paradijs \& Bleeker 1999), and particularly the
chapters by R.~Mewe and D.~Liedahl.

\acknowledgements

We are indebted to D.\ Liedahl for his significant contributions to
this work, and thank H.~Tananbaum for the generous grant of Director's
Discretionary Time which made the {\it Chandra\/} observations possible.
Support for this work was provided in part by NASA Long-Term Space
Astrophysics Program grant S-92654-F and NASA {\it Chandra\/} Guest
Observer grant NAS8-39073. This work was performed under the auspices
of the U.S.\ Department of Energy by University of California Lawrence
Livermore National Laboratory under contract No.\ W-7405-Eng-48.

%%%%%%%%%%%%%%%%%%%%%%%%%%%%%%%%%%%%%%%%%%%%%%%%%%%%%%%%%%%%%%%%%%%%%%%%

%%%%%%%%%%%%%%%%%%%%%%%%%%%%%%%%%%%%%%%%%%%%%%%%%%%%%%%%%%%%%%%%%%%%%%%%

\end{document}